%% file: main.tex
\documentclass[sigplan,nonacm]{acmart}

\usepackage{tikz}
\usepackage{amsmath}
\usepackage{physics}

\usepackage[noend]{algpseudocode}
\usepackage{algorithm}
\usepackage{xspace}
\usepackage[compat=1.1]{yquant}
\usepackage{hyperref}
\usepackage{cleveref}
\usepackage{multirow}
\newcommand{\defn}[1]{\textbf{#1}}

\usepackage{pgfplots}
\usepackage{subcaption}

\pgfplotsset{compat=newest}

\newcommand{\optName} {QALM\xspace}
\newcommand{\quartz}{Quartz\xspace}
\newcommand{\qiskit}{Qiskit\xspace}
\newcommand{\queso}{QUESO\xspace}

\newcommand{\quarl}{Quarl\xspace}

\newcommand{\tket}{t$\ket{\text{ket}}$\xspace}
\newcommand{\voqc}{VOQC\xspace}
\newcommand{\guoq}{GUOQ\xspace}

\usepackage{comment}
\begin{document}
%don't want date printed
\date{}
% make title bold and 14 pt font (Latex default is non-bold, 16 pt)\
%\title{\Large \bf \optName: Invoking Search-Based Optimization to
% Avoid Local Minima  in Rule-Based Quantum Circuit Optimization}
\title{\optName: Escaping Local Minima via Interleaved Exploration and Exploitation in Quantum Circuit Optimization}

%for single author (just remove % characters)
% \author{
%   {\rm Aidan Wagner\footnotemark[1]}\\
%   Carnegie Mellon University
%   \and
%   {\rm Mingkuan Xu\footnotemark[1]}\\
%   Carnegie Mellon University
%   % copy the following lines to add more authors
%   \and
%   {\rm Pengyu Liu\footnotemark[1]}\\
%   Carnegie Mellon University
%   \and
%   {\rm Zhihao Jia}\\
%   Carnegie Mellon University
%   \and
%   {\rm Umut A. Acar}\\
%   Carnegie Mellon University
% } % end author
\author{Aidan Wagner}
\authornote{Contributed equally.}
\affiliation{\institution{Carnegie Mellon University}
\city{Pittsburgh}
\state{PA}
\country{USA}}
\email{aidan.wagner140@gmail.com}

\author{Mingkuan Xu}
\authornotemark[1]
\affiliation{\institution{Carnegie Mellon University}
\city{Pittsburgh}
\state{PA}
\country{USA}}
\email{xumingkuan0721@126.com}

\author{Pengyu Liu}
\authornotemark[1]
\affiliation{\institution{Carnegie Mellon University}
\city{Pittsburgh}
\state{PA}
\country{USA}}
\email{pengyuliu@cmu.edu}

\author{Zhihao Jia}
\affiliation{\institution{Carnegie Mellon University}
\city{Pittsburgh}
\state{PA}
\country{USA}}
\email{zhihao@cmu.edu}

\author{Umut A. Acar}
\affiliation{\institution{Carnegie Mellon University}
\city{Pittsburgh}
\state{PA}
\country{USA}}
\email{umut@cmu.edu}

%\author{\rm ASPLOS Submission \# 2083}

\input{texts/0-abstract.tex}

%\renewcommand{\thefootnote}{\fnsymbol{footnote}}
%\footnotetext[1]{Contributed equally.}

\maketitle
\input{texts/1-introduction.tex}

\input{texts/2-background.tex}

\input{texts/3-method.tex}

\input{texts/4-implementation.tex}

\input{texts/5-eval.tex}

\input{texts/6-related_works.tex}

\input{texts/7-conclusion.tex}

\input{texts/8-ack.tex}

% use the ACM bibliography style
\bibliographystyle{ACM-Reference-Format}
\bibliography{ref.bib}

\end{document}

%% file: texts/0-abstract.tex
\begin{abstract}
  %Two major types of quantum circuit optimizers explored in recent
  % literature are search-based and rule-based approaches. The
  % search-based strategy guarantees that the resulting circuit is
  % optimal with respect to a chosen metric, at the expense of
  % increased optimization time. In contrast, rule-based strategies
  % deliver faster optimization, but lack the optimality guarantee
  % and can become trapped in local minima.
  %This work introduces a hybrid optimization framework that combines
  % the speed of rule-based methods with the exhaustive search
  % capabilities of search-based techniques, enabling the escape from
  % local minima while maintaining efficiency. As a realization of
  % this framework, we propose \optName, an optimizer that achieves
  % both reduced optimization time and improved circuit quality
  % compared to existing methods.
  Quantum circuit optimizers face a fundamental limitation in how
  they tolerate temporary cost increases. At one extreme, greedy
  rule-based optimizers immediately apply any cost-reducing
  transformation, achieving high efficiency but quickly becoming
  trapped in local minima. At the other extreme, search-based
  optimizers accept cost-increasing moves to explore the circuit
  space and escape such minima. However, because search-based
  optimizers cannot determine within a reasonable time budget
  whether a given point is \emph{promising}, that is, whether its
  neighborhood contains a deeper local minimum, they must blindly
  explore higher-cost regions. As a result, escaping the current
  basin to reach a promising point takes exponentially many steps.

  In this work, we show that this limitation can be \emph{overcome}
  with a hybrid framework that interleaves the exhaustive exploration
  capabilities of search algorithms with the efficiency of rule-based
  optimization. We implement this framework as \optName{}, a novel
  optimizer designed to escape local minima without incurring the
  runtime penalties of pure search. Crucially, our results demonstrate
  that \optName{} does not merely strike a balance; it outperforms
  existing rule-based and search-based optimizers in circuit reduction
  rates while operating with the computational efficiency of rule-based
  systems. In a comprehensive evaluation across 248 circuits,
  \optName{} matches or exceeds the fidelity of the strongest baseline
  on $83.9\%$ of these circuits, given the same time budget.
\end{abstract}

%% file: texts/1-introduction.tex
\section{Introduction}
Quantum computers have shown potential to solve problems that are
intractable for classical machines. Proposed and demonstrated
applications span fundamental quantum
physics~\cite{feynman2018simulating}, computational
chemistry~\cite{babbush2015chemical},
cryptography~\cite{shor1994algorithms,bennett2014quantum}, machine
learning~\cite{zhang2016quantum,schuld2019quantum}, and
finance~\cite{orus2019quantum}. Regardless of whether a machine is a
near-term noisy device or a future fault-tolerant quantum computer,
the resource cost of a circuit (gate count, two-qubit gate count,
circuit depth, $T$ count, or $T$ depth) directly determines runtime,
error accumulation, and, for error-corrected systems, the overhead
required for logical operations. Consequently, circuit
optimization---the task of producing equivalent implementations that
minimize those
costs---is central across the entire quantum computing stack and lifecycle.

% Circuit optimization (or compilation) aims to transform a given
% logical circuit into an equivalent implementation that uses fewer
% gates, less depth, or hardware-friendlier primitives.
A wide range of hardware-independent and hardware-aware techniques
has been developed for circuit optimization, spanning algebraic and
rewrite-based simplification, peephole passes, constraint-aware qubit
routing, and automated diagrammatic approaches such as the
ZX-calculus~\cite{quartz-2022,queso-2023,nam2018automated,hietala2021verified,cowtan2019qubit,kissinger2020Pyzx}.
Together, these techniques form the modern quantum-compiler toolchain
and are widely surveyed in the recent
literature~\cite{karuppasamy2025comprehensive}.

%\mingkuan{x-axis optimization time, y-axis performance; so upper
% left is better; don't show vanilla combination}
% \begin{figure}
%   \centering
%   \includegraphics[width=\linewidth]{figure/tradeoff.pdf}
%   \caption{The trade-off space between performance and optimization
%     time. \optName{} breaks this barrier by interleaving rule-based
%     and search-based strategies. It uniquely occupies the
%     high-performance, low-latency quadrant, a region previously
%     inaccessible to existing techniques.
%     %integrates rule-based and
%     % search-based methods, outperforming both individual techniques
%     % while maintaining significantly higher speeds than search-based
%     %optimizers alone.
%   }
%   \label{fig:tradeoff}
% \end{figure}
\input{figtex/fig1.tex}

Existing circuit optimizers face fundamental limitations
(\Cref{fig:fig1}): rule-based optimizers greedily descend into the
nearest local minimum and terminate, resulting in fast optimization
but low quality.

To escape one local minimum and reach a better one, search-based
optimizers explore neighboring points exhaustively. However, among
all the explored points, it is difficult for a search-based
optimizer to distinguish a \emph{promising point}, one that is
likely to lead to a better local minimum, from one that merely
undoes previous progress. To make this distinction, existing
search-based optimizers must continue the exhaustive search for
exponentially many steps until the promising point actually
descends into a better local minimum.

This issue fundamentally limits the performance of search-based
optimizers. Prior works attempt to mitigate it, including
cost-based backtracking search~\cite{quartz-2022, queso-2023},
Monte Carlo tree search~\cite{zhou2022quantum}, and reinforcement
learning~\cite{li2024quarl}, but most of them only trade one point
on the speed--quality frontier for another rather than escaping the
frontier itself.
% \mingkuan{In related work, talk about GUOQ proposes the fast and slow
%   idea, but the ``fast'' is still very slow, and it focuses on
% approximate optimization}

% This paper argues that the trade-off can be improved by dynamically
% balancing exploration and exploitation during the optimization
% process. Intuitively, certain regions of the ``equivalence
% landscape'' benefit from broad exploration (to discover promising and
% diverse rewrites), while others reward deep exploitation (to fully
% compress a promising subcircuit). We propose viewing circuit
% optimization as a sequential decision process in which the optimizer
% adaptively chooses between fast, local passes and slower, more global
% search moves. By combining these modes, we show that it is possible
% to reach near-best compression while using substantially less
% wall-clock time than applying full global search everywhere.

This paper shows that this limitation can be \emph{overcome}:
\emph{good optimization quality} and \emph{efficient runtime} are
not inherently in tension. Rather than simply balancing exploration and
exploitation, we \emph{interleave} search-based and rule-based
optimization in a unified control loop. As illustrated in
\Cref{fig:fig1}, exploit phases descend to a local minimum as
quickly as a rule-based optimizer, while interleaved explore
phases uncover candidate promising points via search; the
following exploit phase then tests each candidate by descending
from it, keeping the one that reaches the deepest local minimum. This
design identifies the truly promising
point within a single exploit phase, sidestepping the exponential
wait incurred by pure search.
%  Intuitively, certain regions
% of the ``equivalence
% landscape'' benefit from broad exploration (to discover promising and
% diverse rewrites), while others reward deep exploitation (to fully
% compress a promising subcircuit). We formulate circuit optimization
% as a sequential decision process in which the optimizer interleaves
% fast, local rule-based passes with slower, more global search moves.
% Through this synergistic combination, we
% show that it is possible to beat existing optimizers while using
% substantially less wall-clock time than applying full global search
% everywhere.

We implement this framework as \optName{} (\Cref{sec:method}), a
hybrid optimizer that dynamically interleaves exploration and
exploitation to escape local minima.
%As shown
% in Section \ref{sec:method}, \optName uses a strategic combination of
% effective and rapid optimization passes to both navigate quickly
% through the loss landscape while escaping potential local minima.

Our evaluation on 248 circuits shows that \optName{} significantly
outperforms all existing rule-based and search-based optimizers.
Given the same time budget (one hour), \optName{} achieves 5.97\% more gate
count reduction than the strongest baseline, and even when given
only one minute, it still achieves 4.98\% more reduction than the
strongest baseline.

\input{figtex/eval_win_rate.tex}

This advantage also holds on a per-circuit basis, as shown in
\Cref{fig:win_rate}. Compared with \queso{} (in both two-qubit
and total gate optimization modes), \voqc{}, \quartz{}, and \guoq{}
(with total gate optimization mode), \optName{} wins or ties on
nearly every circuit across all three
metrics. Although the two-qubit gate reduction is on par with \guoq{}
(with two-qubit gate optimization mode), \optName{}'s significantly
lower total gate count still leads to higher or equal fidelity on 83.9\% of circuits.

%% file: figtex/fig1.tex
\begin{figure}[t]
  \centering
  \includegraphics[width=\columnwidth]{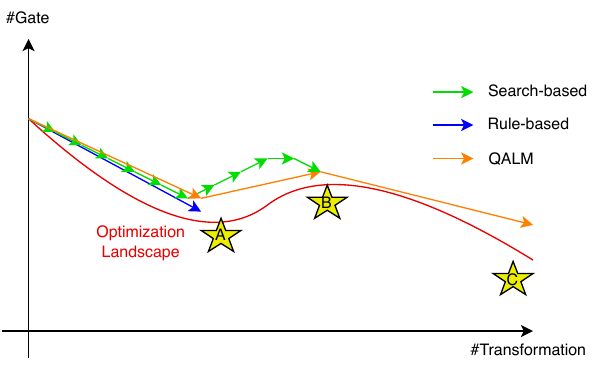}
  \caption{Comparison of optimization trajectories across three
    strategies: rule-based (pure exploitation), search-based (pure
    exploration), and \optName{} (interleaved exploration and
    exploitation). Rule-based optimization greedily descends to a
    local minimum in a single step and terminates (shown as $A$). Search-based
    optimization reaches the same local minimum more slowly, then
    continues to explore and discovers a candidate region; however,
    because this new region has a higher cost (shown as $B$), and the
    search-based method cannot determine whether this region is truly promising,
    escaping the current
    basin requires exponentially many search steps. \optName{}
    interleaves the two phases: the exploit phase quickly reaches the
    local minimum, and the subsequent explore phase discovers the
    promising region just as the search-based method does. Crucially,
    the next exploit phase immediately capitalizes on this discovery,
    descending to a new solution (shown as $C$) that surpasses the
    original local
    minimum. Through this interleaving, \optName{} finds much better
    solutions than both methods with speed comparable to rule-based
    optimization.
  }
  \label{fig:fig1}
\end{figure}

%% file: figtex/eval_win_rate.tex
\begin{figure}[t]
  \centering
  \includegraphics[width=\columnwidth]{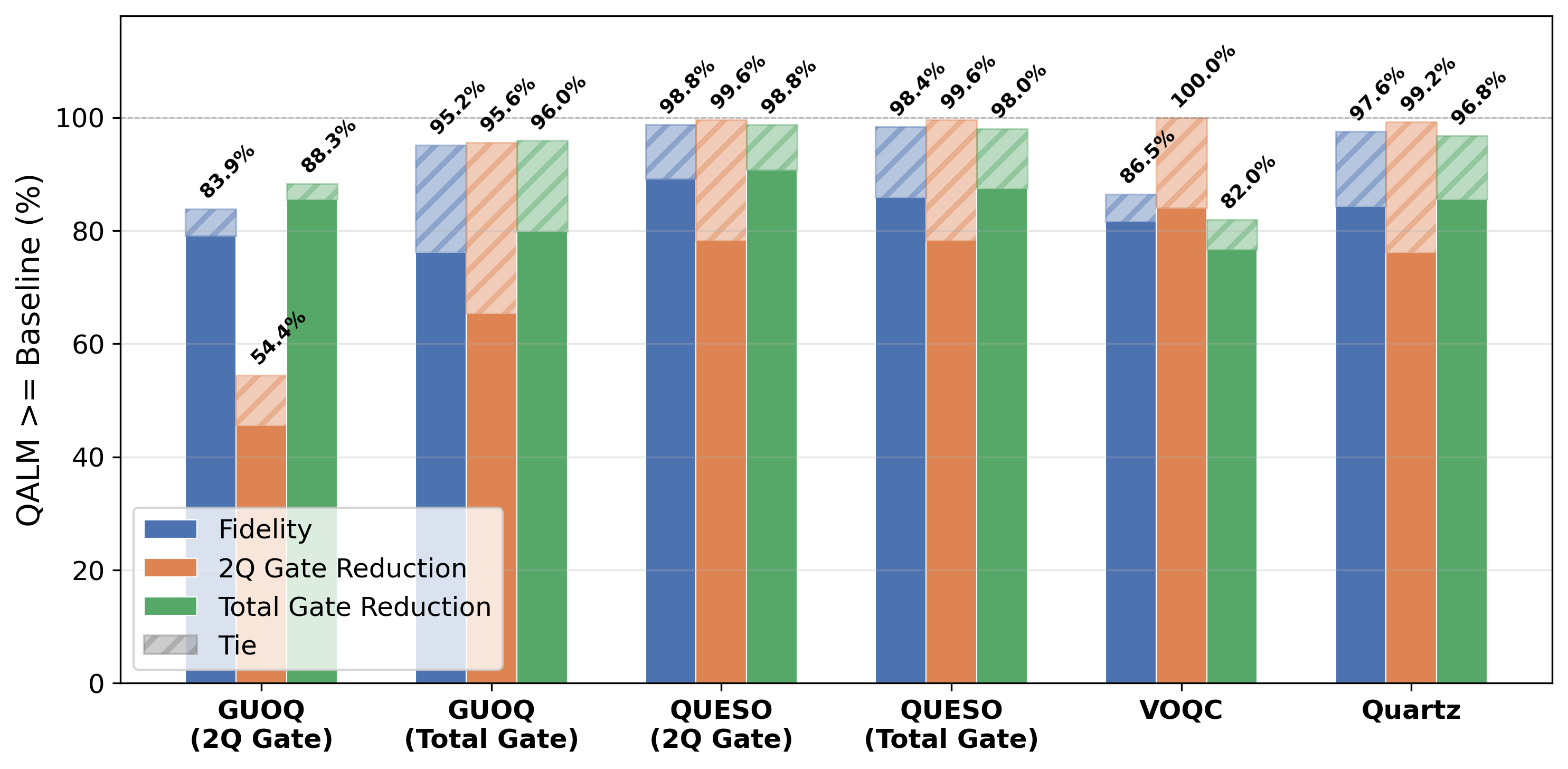}
  \caption{Percentage of circuits on which \optName{} matches or
    exceeds each baseline across fidelity, $CX$ gate reduction, and
    total gate reduction. \guoq{} and \queso{} are each shown under
  two configurations: one optimizing for $CX$ count and one for total gate count.}
  \label{fig:win_rate}
\end{figure}

%% file: texts/2-background.tex
\section{Background}
\label{sec:background}

In this section, we introduce quantum circuit optimization as a
\emph{graph transformation problem}. While we provide essential
background on quantum circuits, readers familiar with compiler
optimization will recognize familiar themes: quantum circuit
optimization closely parallels classical compiler passes such as
peephole optimization and instruction scheduling.

\subsection{Quantum Computation}

Classical computers operate on \defn{bits}, each storing either 0 or
1. In quantum computation, the basic unit is a \defn{qubit}, which can
exist in a \emph{superposition} of both 0 and 1 states
simultaneously~\cite{10.5555/1972505}. This is analogous to how a
probability distribution can assign weights to multiple outcomes.

The key insight for understanding circuit optimization is that an
$n$-qubit system has $2^n$ possible basis states. This exponential
growth is what makes quantum computers potentially powerful, but also
what makes global circuit optimization intractable.
% verifying
% equivalence between two circuits requires reasoning about an
% exponentially large state space.

\subsection{Quantum Circuits}

Quantum circuits provide a compact representation of quantum
algorithms, analogous to how dataflow graphs represent classical
computations. A quantum circuit can be viewed as a \emph{directed
acyclic graph (DAG)}: each qubit corresponds to a ``wire,'' and gates
are nodes that transform the states flowing through these
wires~\cite{10.5555/1972505}. Two circuits are considered
equivalent if they produce the same output for all possible
inputs, similar to how two code sequences are equivalent if they
compute the same function.

This graph-based view is central to circuit optimization: just as
classical compilers apply rewrite rules to transform instruction
sequences, quantum circuit optimizers apply \emph{equivalence-preserving
graph transformations} to reduce circuit cost.

A gate set defines the allowed gate types in a circuit.
Throughout this paper, we use the standard Nam
gate set~\cite{nam2018automated}: single-qubit gates ($X$, $H$, $R_z$)
and the two-qubit controlled-NOT gate ($CX$). This universal gate set
is widely adopted by quantum circuit optimizers.

\subsection{Quantum Circuit Optimization}

Quantum circuit optimization is fundamentally a \emph{graph rewriting
problem}: given an input circuit (DAG), find an equivalent circuit
with minimum cost. In this paper, we primarily focus on minimizing the
total gate count. This metric is hardware-independent, making it
suitable for benchmarking across different quantum architectures, and
is widely adopted by prior work~\cite{nam2018automated,quartz-2022,li2024quarl},
enabling direct comparison. We also evaluate two-qubit ($CX$) gate
count reduction in \Cref{sec:eval}.

This problem is provably hard: determining whether two circuits are
equivalent is QMA-complete~\cite{janzing2003identity}, a quantum
generalization of NP-completeness. In practice, this means that no
polynomial-time algorithm can globally optimize arbitrary circuits.
This situation mirrors classical compiler optimization, where problems
like optimal register allocation and instruction scheduling are also
NP-hard~\cite{alfred2007compilers}, forcing compilers to rely on heuristics.

Existing quantum circuit optimizers fall into two main categories:
\emph{rule-based} and \emph{search-based} strategies.

\subsubsection{Rule-Based Optimization}

\begin{figure}[h]
  \centering
  \begin{minipage}{0.9\linewidth}
    \centering
    \begin{tikzpicture}
      \begin{yquant}
        qubit {$q_0$} q0;
        qubit {$q_1$} q1;

        box {$R_z(\theta_1)$} q1;
        cnot q1 | q0;
        box {$R_z(\theta_2)$} q1;
        cnot q1 | q0;
        box {$R_z(\theta_3)$} q1;
      \end{yquant}
    \end{tikzpicture}
  \end{minipage}

  \vspace{0.3cm}
  $\Downarrow$
  \vspace{0.3cm}

  \begin{minipage}{0.9\linewidth}
    \centering
    \begin{tikzpicture}
      \begin{yquant}
        qubit {$q_0$} q0;
        qubit {$q_1$} q1;

        cnot q1 | q0;
        box {$R_z(\theta_2)$} q1;
        cnot q1 | q0;
        box {$R_z(\theta_1 + \theta_3)$} q1;
      \end{yquant}
    \end{tikzpicture}
  \end{minipage}

  \caption{Example of an optimization rule for a gate pattern with
  multiple $CX$ and $R_z$ gates.}
  \label{fig:RZandCNOTRule}
\end{figure}
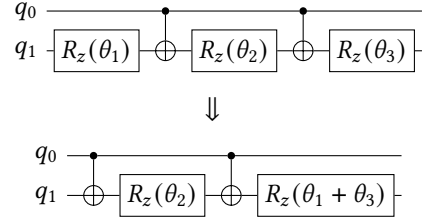

Rule-based optimization applies a predefined set of local
transformation rules, analogous to \emph{peephole optimization} in
classical compilers. Each rule specifies a pattern of gates that can
be replaced with an equivalent but cheaper sequence. One simple
example is that two adjacent Hadamard gates acting on the same
qubit can be reduced to the identity operation, as the Hadamard
gate is its own inverse. This is illustrated in
\Cref{fig:HadRule}. For a more complex reduction rule, see
\Cref{fig:RZandCNOTRule}, in which we reduce the total number of
rotation gates applied to the bottom qubit. These are just two of
the many rules that can be applied to reduce the total gate count.

\begin{figure}[h]
  \centering
  \begin{minipage}{0.35\linewidth}
    \centering
    \begin{tikzpicture}
      \begin{yquant}
        qubit {$q_0$} q;
        h q;
        h q;
      \end{yquant}
    \end{tikzpicture}
  \end{minipage}
  $\Rightarrow$
  \begin{minipage}{0.35\linewidth}
    \centering
    \begin{tikzpicture}
      \begin{yquant}
        qubit {$q_0$} q;
        box {$I$} q;
      \end{yquant}
    \end{tikzpicture}
  \end{minipage}

  \caption{Hadamard self-inverse property: $HH = I$, where $I$ denotes the identity gate, representing no operation on the qubit.}
  \label{fig:HadRule}
\end{figure}

Rule-based optimizers repeat these substitutions heuristically
with the goal of reducing the total gate count of the quantum circuit.
However, rule-based optimization is inherently greedy: substitutions
are made without considering how these changes may impact future
opportunities for additional rule applications. This can cause the
optimizer to become stuck in a local minimum,
unable to explore the optimization landscape at a large enough scale
to find better solutions. One possible strategy for remedying this
concern is to use a search-based approach.

\subsubsection{Search-Based Optimization}

Like rule-based approaches, search-based optimization
considers a set of possible transformations that preserve
the equivalence of the circuit. However, unlike rule-based approaches
that greedily apply beneficial transformations, search-based
optimizers systematically explore the space of equivalent circuits,
tolerating transformations that do not reduce---or even
increase---circuit size. This allows them to escape local minima by
exploring a much larger portion of the optimization landscape.
As a result, search-based optimizers usually require longer runtimes
to explore a reasonably large portion of the search space, but they
typically achieve better optimization results than rule-based
optimizers given sufficient time.

To illustrate the advantage of search-based optimization, consider
the example in \Cref{fig:search_advantage}. In the first step of
this example, two $CX$ gates are transformed into three $CX$ gates. While
this individual transformation
increases the total gate count (and thus might be rejected by a
greedy rule-based approach), it alters the circuit
structure in a way that enables subsequent optimizations, such as $CX$
cancellations, that were not previously accessible. This highlights the
ability of search-based methods to traverse non-improving steps to
reach a better local optimum.

\input{figtex/search_advantage.tex}

%% file: figtex/search_advantage.tex
\begin{figure}[t]
  \centering
  \includegraphics[width=0.9\columnwidth]{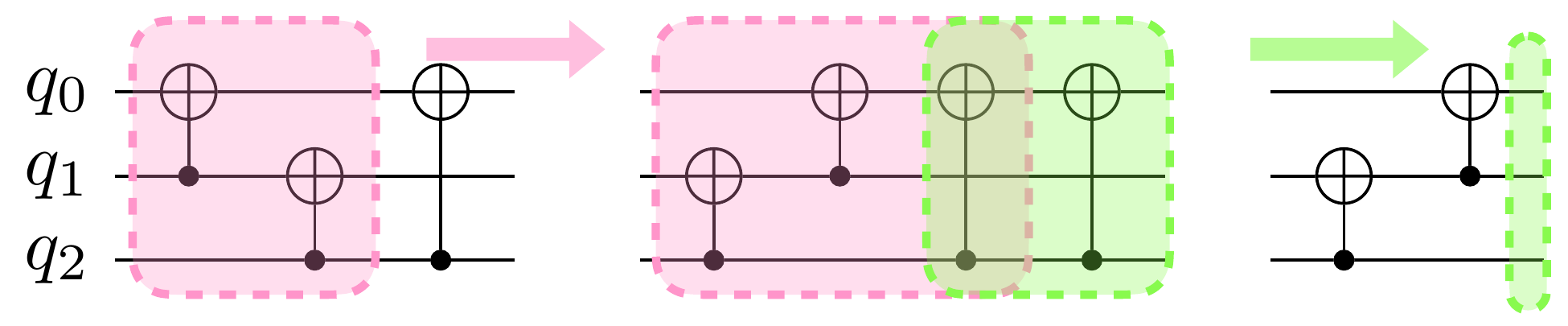}
  \caption{Search-based optimization example. The first transformation
    increases gate count ($3 \rightarrow 4$), but enables subsequent
  $CX$ cancellation ($4 \rightarrow 2$).}
  \label{fig:search_advantage}
\end{figure}

%% file: texts/3-method.tex
\section{Method}
\label{sec:method}
Current approaches to quantum circuit optimization typically fall
into one of two distinct paradigms: rule-based optimization and
search-based optimization. We propose that these are not mutually
exclusive methodologies, but rather two extremes of a single,
continuous optimization spectrum. In this section, we introduce a
unified framework that parameterizes the trade-off between
exploration (search) and exploitation (rules). We define a parameter
$k$, representing the depth of search exploration permitted before
applying a rule-based reduction. By tuning $k$, we can recover
existing distinct methodologies or interpolate between them to create
a hybrid approach, the \optName{} algorithm, that captures the
efficiency of rule-based systems and the comprehensiveness of
search-based systems.

\subsection{Optimization Spectrum}
We model the optimization process as a traversal through a circuit
space, seeking to minimize a cost function (e.g., total gate count or
two-qubit gate count).

Rule-based optimization (Exploitation): Optimizers like
\qiskit~\cite{qiskit2024}, \voqc~\cite{hietala2021verified}, and
\tket~\cite{tket} rely on pattern matching. They are computationally
efficient but ``greedy'', immediately applying transformations when a
pattern is matched to reduce the cost. This often leads to being
trapped in local minima.

Search-based optimization (Exploration): Optimizers like
\quartz~\cite{quartz-2022}, \queso~\cite{queso-2023}, and
\guoq~\cite{xu2024optimizing} utilize cost-increasing transformations
to explore the search space exhaustively. They effectively escape
local minima but suffer from exponential time complexity as the
search depth increases.

We unify these approaches by introducing an exploration depth
parameter, denoted as $k$. This parameter dictates the number of
transformation steps the optimizer takes to expand the search
frontier before applying a rule-based reduction pass to collapse the frontier.
The spectrum can be defined as follows:
\begin{itemize}
  \item $k = 0$ (Pure rule-based): The optimizer performs zero search
    steps before applying rules. This is equivalent to rule-based optimizers.
  \item $k \to \infty$ (Pure search-based): The optimizer performs
    infinite (or exhaustive) search steps. This is equivalent to
    search-based optimizers.
  \item $0 < k < \infty$ (Hybrid): This is the domain of our proposed
    method. By setting a finite, non-zero $k$, we allow the optimizer
    to explore the neighborhood of the current circuit before using
    the rule-based optimizer to jump to a deeper local minimum.
\end{itemize}

\subsection{The \optName{} Algorithm}
\label{sec:main_algo}
Our algorithm, \optName{}, operates within the hybrid region of this
spectrum. It utilizes a multi-pass strategy that progressively
increases the exploration depth $k$. This allows the optimizer to
harvest ``low-hanging fruit'' with fast, shallow searches before
committing computational resources to deeper exploration.
Since the branching factor and effective search depth of general
search-based optimization techniques scale with the size of the
circuit, employing rule-based reduction as a preprocessing step
significantly shrinks the search space. This ensures that the
computationally expensive exploration (high $k$) is performed on a
compacted circuit topology, preventing the search algorithm from
wasting its limited depth budget on trivial cancellations that could
have been resolved greedily.
% (another version 2) We impose this ordering because search-based
% optimization is sensitive to initial conditions. By applying
% rule-based optimization first, we effectively lower the starting
% point of the search traversal into a deeper local basin. This
% prevents the search algorithm from expending its initial
% computational effort merely descending the slope of a trivial local
% minimum, and instead allows the exploration to immediately probe
% the boundaries of the local basin for escape paths to globally
% superior solutions.

\input{figtex/top_level_algo.tex}

\Cref{alg:qalm} details the \optName{} execution flow. \optName{}
runs iteratively in passes, with the exploration depth $k$ increasing
from one until the time limit is reached.
In each pass, \optName{} maintains a priority queue for circuit
candidates (\Cref{alg:qalm} line~\ref{alg1:priority_queue}), similar
to search-based optimizers.
%\mingkuan{Add Npool and Nbranch to the discussion, (in our
% preliminary evaluation, we found N=1 is the best. Evaluation
% doesn't need to be in the evaluation section. These parameters
% enable us to trade-off between depth-first and breadth-first
% search. Leaving them allows us to control the search space
% depending on the hyperparameters).  Trade-off depending on search
% time? We can just show results in text. Ideally we can have the curve.}

In each iteration of the search, \optName{} creates a circuit
candidate pool according to the schedule $N_{\rm{pool}}^{(k)}$ (lines
\ref{alg1:pool_begin} to \ref{alg1:pool_end}) for parallel exploration.
For each circuit in the candidate pool, \optName{} branches
$N_{\rm{branch}}^{(k)}$ times (line~\ref{alg1:branch}) and explores
for $k$ steps per branch (line~\ref{alg1:explore_step}). Each of the
$k$ steps is a search step in the search-based optimizer
(line~\ref{alg2:explore_step} in \Cref{alg:explore}, \Cref{sec:search}).
After the $k$ steps, \optName{} calls a rule-based optimizer for an
exploitation step (line~\ref{alg1:exploit}, \Cref{sec:roqc}).
\optName{} then enqueues all $N_{\rm{branch}}^{(k)}$ new circuits
after the exploitation step if they have not been visited already.

While $k$ governs the balance between exploration and exploitation,
the schedules $N_{\rm{pool}}^{(k)}$ and $N_{\rm{branch}}^{(k)}$
determine the exploration \emph{strategy}.
%Specifically, setting $N_{\rm{pool}}^{(k)} = N_{\rm{branch}}^{(k)} =
% 1$ reduces to a cost-based depth-first search (DFS), whereas
% $N_{\rm{pool}}^{(k)}, N_{\rm{branch}}^{(k)} \to \infty$ approaches
% a cost-based breadth-first search (BFS). This allows us to
% interpolate between these two regimes. We focus on the DFS
% configuration for our primary evaluation and analyze the impact of
% search width in \Cref{sec:eval_explore}.
Specifically, setting $N_{\rm{pool}}^{(k)} = 1$ reduces to a
best-first search, whereas higher $N_{\rm{pool}}^{(k)}$ and
$N_{\rm{branch}}^{(k)}$ produce a wider beam search. The schedule
formulation makes the framework general: in particular, the greedy
mode of \Cref{sec:greedy} is a different schedule for $k \le 2$ that
effectively forces $N_{\rm{pool}} = N_{\rm{branch}} = 1$ and rejects
cost-increasing moves. For our main search loop, we use the constant
schedule $N_{\rm{pool}}^{(k)} = 1$ and $N_{\rm{branch}}^{(k)} = 3$
and analyze their impact in \Cref{sec:eval_explore}.

%% file: figtex/top_level_algo.tex
\begin{algorithm}[ht]
\caption{\optName{} Algorithm.}\label{alg:qalm}
\begin{algorithmic}[1] % The number tells where the line numbering should start
\State {\bf Input:} Quantum circuit $C_\text{in}$, cost model $\Call{Cost}{\cdot}$, schedules $N_{\text{pool}}^{(k)}, N_{\text{branch}}^{(k)}$, and a time limit.
\State {\bf Output:} An optimized circuit $C_\text{out}$.
\State $C_\text{out} \gets C_\text{in}$
%\State $t_{\text{start}} \gets \text{CurrentTime}()$
\For{$k = 1, 2, \dots$ while time limit not exceeded}
    %\If{$\text{CurrentTime}() - t_{\text{start}} > T_{\max}$}
    %    \State \textbf{break}
    %\EndIf
    \State $C_\text{out} \gets \Call{Optimize}{C_\text{out}, k}$
\EndFor
\State \Return $C_\text{out}$
\State
\Function{Optimize}{$C_\text{in}, k$}
\State {\em // $\mathcal{Q}$ is a priority queue of circuits sorted by their $\Call{Cost}{\cdot}$.}
\State $\mathcal{Q} \gets \{C_\text{in}\}$ \label{alg1:priority_queue} %\Comment{$\mathcal{Q}$ is a priority queue of circuits sorted by their $\Call{Cost}{\cdot}$.}
\State $\mathcal{V} \gets \{C_\text{in}\}$ \Comment{$\mathcal{V}$ is the set of visited circuits}
\State $C_\text{best} \gets C_\text{in}$
\While{$\mathcal{Q} \neq \emptyset$ and time limit not exceeded} \label{alg1:main_queue_loop}
    \State {\em // Make an initial circuit candidate pool of size $N_{\rm{pool}}^{(k)}$ from the priority queue.} \label{alg1:pool_begin}
    \State $\mathcal{P} \gets \emptyset$
    \For {$i \gets 1 \textbf{ to } N_{\text{pool}}^{(k)}$ and $\mathcal{Q} \neq \emptyset$}
        \State $C \gets \mathcal{Q}$.dequeue()
        \If{$\Call{Cost}{C} < \Call{Cost}{C_{best}}$}
            \State $C_\text{best} \gets C$
        \EndIf
        \State $\mathcal{P} \leftarrow \mathcal{P} \cup \{C\}$
    \EndFor \label{alg1:pool_end}
    \State {\em // Branch, explore, and exploit.}
    \For {each circuit $C \in \mathcal{P}$}
        \For {$i \gets 1 \textbf{ to } N_{\text{branch}}^{(k)}$}  \label{alg1:branch}\Comment{\!Branch $N_{\text{branch}}^{(k)}$ times}
            %\For {$j \gets 1 \textbf{ to } k$} \label{alg1:explore_loop} \Comment{Explore for $k$ steps}
            \State $C \gets \Call{Explore}{C, k}$  \Comment{Explore $k$ steps} \label{alg1:explore_step}
            %    \State $\mathcal{Q}$.enqueue($C$)
            %\EndFor
            \State $C \gets \Call{Exploit}{C}$\label{alg1:exploit} \Comment{Exploit for one step} 
            \If {$C \notin \mathcal{V}$}
                \State $\mathcal{Q}$.enqueue($C$)
                \State $\mathcal{V} \leftarrow \mathcal{V} \cup \{C\}$
            \EndIf
        \EndFor
    \EndFor
\EndWhile
\State \Return $C_\text{best}$
\EndFunction
\end{algorithmic}
\end{algorithm}

%% file: texts/4-implementation.tex
\section{Implementation}

This section describes the implementation of \optName{}, which
integrates a rule-based optimizer and a search-based optimizer under
a unified control loop.
We first describe our rule-based component, ROQC (\Cref{sec:roqc}),
which synthesizes techniques from several prior rule-based
optimizers~\cite{hietala2021verified,nam2018automated,chen2025phasepoly}.
We then present the search-based component (\Cref{sec:search}),
which leverages transformation rules from
Quartz~\cite{quartz-2022}.
Finally, we discuss a greedy mode that accelerates termination for
small $k$ (\Cref{sec:greedy}).
% Finally, we present the implementation
% of the QALM algorithm
% (\S\ref{sec:qalm}) that orchestrates the interleaving of these two
% phases, and discuss several search strategy variants (\S\ref{sec:variants}).

\subsection{ROQC: Rule-based Optimizer for Quantum Circuits}
\label{sec:roqc}

ROQC is our rule-based optimizer that implements all optimization
routines from VOQC~\cite{hietala2021verified}, including Not
propagation, Hadamard reduction, single-qubit cancellation, two-qubit
gate cancellation, and
rotation merging.
%To further improve optimization quality and speed,
%we draw inspiration from the floating $R_z$ gate technique proposed by Nam et
%al.~\cite{nam2018automated}, which computes all possible locations to
%which $R_z$ gates can be moved.
%Rather than computing all such
%locations, we implement \textit{bidirectional rotation merging}, which
%pushes $R_z$ gates toward the beginning, cancels adjacent $CX$
% gates, and then pushes $R_z$ gates toward the end
%of the circuit. %\pengyu{Is this accurate?}
%This approach captures many additional optimization
%opportunities missed by the original rotation merging routine, while
%requiring only $2\times$ time overhead compared to the $O(n)$ multiplicative
%overhead of the floating-$R_z$ approach, where $n$ is the number of
%gates.
%NOTE: \aidan{I think the two way merging adds an additional O(n), is
% that not what floating Rz does?}
% This complexity claim should be verified and ideally supported with a
% micro-benchmark or a brief complexity analysis.
To further improve optimization quality and speed,
we extend rotation merging with the tagging technique from
PhasePoly~\cite{chen2025phasepoly}, which assigns a tag to
each $R_z$ gate so that gates with the same tag can be merged. This
finds more merging opportunities than the previous $R_z$-$CX$ block approach.
% that synthesizes and extends
% techniques from several prior works, including
% VOQC~\cite{hietala2021verified}, Nam et al.~\cite{nam2018automated}, and
% PhasePoly~\cite{chen2025phasepoly}. ROQC incorporates five
% optimization routines:
% \mingkuan{Shall we cite them individually? I feel like we want to
%   describe a bit more about tag-based rotation merging, but saying
%   "Please refer to PhasePoly..." sounds abrupt, so I want to cite the
% works for each routine instead.}

\subsection{Search-Based Optimization}
\label{sec:search}

For the search-based component, we leverage the transformation rules
from Quartz~\cite{quartz-2022}. Specifically, we use the
$(5,3)$-complete circuit transformation set (ECC set), which generates
equivalence-preserving transformations involving at most 3 qubits and
at most 5 gates. The transformation set $\mathcal{T}$ in
\Cref{alg:explore,alg:explore_greedy} is
pre-computed and verified for correctness, allowing \optName{} to
apply these transformations efficiently during the search phase.
%In the search phase, we apply circuit transformations uniformly at random.
%\optName{} selects the transformation locations uniformly at
%random, which means that consecutive transformations can target distant parts
%of the circuit without interaction.

During exploration, \optName{} biases transformation selection toward recently
modified regions. This encourages the discovery of optimization
sequences in which one transformation enables another, ensuring that
the $k$ exploration steps cannot be easily obtained by performing one
exploration step $k$ times independently.

Similar to \quartz~\cite{quartz-2022}, to prevent the exploration
from consuming too much memory, whenever the priority queue of
\Cref{alg:qalm} contains more than 2,000 circuits, we prune it and
keep only the top 1,000 circuits. In addition, to save memory, we
implement the set of visited circuits $\mathcal{V}$ as a hash table,
and only store the hash value instead of the entire circuit. Our
preliminary experimentation suggested that this pruning does not
affect the results. %\aidan{Were these experiments just from Quartz
% or did we do them for QALM too?}
% \mingkuan{We do them for QALM too}
%NOTE: "Preliminary experimentation suggested it does not affect results" is too
% informal for a paper; consider adding a brief quantitative comparison (e.g.,
% gate counts with and without pruning on a few representative circuits) or
% replacing with a principled argument (e.g., citing similar
% thresholds in \quartz).

\input{figtex/explore_random.tex}

% \subsection{The QALM Algorithm}
% \label{sec:qalm}

% QALM orchestrates rule-based and search-based optimization through an
% interleaved control loop (\Cref{alg:qalm}), using the transformation
% rules from Quartz~\cite{x+quartz-2022} for the search phase. The
% inner loop is governed by three parameters:
% $N_\text{pool}^{(k)}$, the number of circuits pulled from the priority queue
% in each iteration;
% % \mingkuan{They are actually
% % schedules/series, can vary with the epoch $k$, as in algorithm 1}
% $N_\text{branch}^{(k)}$, the number of variant circuits generated from each
% circuit via random transformations; and
% $k$, the number of consecutive search transformations applied before
% invoking rule-based optimization.

% An outer loop adjusts these parameters to favor exploitation early in
% the optimization process, when the circuit is easier to optimize, and
% shifts toward exploration in later stages when further improvements
% become harder to find.
% (i.e., the exploration depth $k$). \mingkuan{If $N_\text{steps}$ is
% never used anywhere, let's just use $k$}

% \subsection{Search Strategy Variants}
% \label{sec:variants}

% The search space of equivalent circuits grows exponentially with
% the circuit size, making exhaustive exploration infeasible. Moreover,
% different stages of optimization may benefit from different
% strategies: early on, greedy search can quickly harvest obvious
% improvements, while later stages may require deeper exploration to
% escape local minima. To address this, QALM supports several
% configurable strategy variants.

\input{figtex/explore_greedy.tex}
\subsection{Greedy Mode}
\label{sec:greedy}
In preliminary experiments, we observed that many circuits contain
``low-hanging fruit''---optimizations discoverable with minimal
search depth. To exploit this, we introduce a \emph{greedy mode} for
the two initial passes with $k=1$ and $k=2$, terminating each
branch early once an improvement is found. This allows \optName{} to
quickly harvest obvious optimizations before committing to deeper exploration.

Specifically, similar to setting $N_\text{pool}$ and
$N_\text{branch}$ to 1, we maintain only one
circuit candidate, and only accept the circuit
transformation if it (after the exploitation step) reduces the
circuit cost, by replacing $\Call{Optimize}{}$ in \Cref{alg:qalm}
with $\Call{OptimizeGreedy}{}$ in \Cref{alg:explore_greedy}.
Instead of looping until a fixed time limit, as shown in
\Cref{alg:explore_greedy} line~\ref{alg:explore_greedy:transformation}, we
only perform the exploration for one pass for each circuit transformation.
In the exploration step, instead of applying a transformation
uniformly at random, we start from the position where we last
applied a circuit transformation to find the next available position
(\Cref{alg:explore_greedy} line~\ref{alg:explore_greedy:wrap_around}),
making the pass more deterministic as the greedy mode is fast and we
can afford to loop over all transformation-position combinations for
one pass. For $k \ge 2$ during the greedy pass, the function
$\Call{GreedyStep}{}$ does not go to the exploitation phase
immediately, and we explore all possible transformations from the
modified indices recursively at
line~\ref{alg:explore_greedy:recursive}. After $k - 1$ levels of
recursion, we run the exploitation step, and stop the recursion chain
to move on to the next iteration if the circuit cost is reduced after
the $k$ transformations and the exploitation step.

We evaluate the empirical benefit of this greedy mode in \Cref{sec:eval_greedy}.

% Not used in final QALM
% \subsubsection{Transformation Filtering}

% The transformation provided by Quartz considers all transformations
% under a fixed qubit and gate count limit,
% and includes both cost-decreasing and cost-increasing transformations.
% (i.e., gate count when we optimize for this
% metric). While cost-increasing transformations
% enable exploration beyond local minima, they also expand the search
% space. We provide an option to \emph{filter out cost-increasing
% transformations}, which restricts the search to non-increasing moves
% only. This variant trades exploration capability for faster
% convergence.

%% file: figtex/explore_random.tex
\begin{algorithm}[t]
\caption{The exploration step.}\label{alg:explore}
\begin{algorithmic}[1]
%\State {\bf Global Input:} Circuit transformations $\mathcal{T}$.
\Require Circuit transformations $\mathcal{T}$.
\Function{Explore}{$C, k$}
\State $S \gets C$ \Comment{Search from all indices in the first step, then from modified indices from the previous step in the current iteration}
\For {$j \gets 1 \textbf{ to } k$}
    \While{\textbf{true}}
        \State transformation $t \gets \Call{SampleUniform}{\mathcal{T}}$
        \State position $c \gets \Call{SampleUniform}{S}$
        \If{$t$ is applicable at $c$}
            \State $(C, S) \gets \Call{Apply}{C, c, t}$ \Comment{Returns new circuit and modified indices} \label{alg2:explore_step}
            \State \textbf{break}
        \EndIf
    \EndWhile
\EndFor
\State \Return $C$
\EndFunction
\end{algorithmic}
\end{algorithm}

%% file: figtex/explore_greedy.tex
\begin{algorithm}[ht]
\caption{The greedy mode ($k \le 2$).}\label{alg:explore_greedy}
\begin{algorithmic}[1]
\Require Circuit transformations $\mathcal{T}$.
\Function{OptimizeGreedy}{$C_\text{in}, k$}
\State $C_\text{best} \gets C_\text{in}$
\For{$t \in \mathcal{T}$} \label{alg:explore_greedy:transformation}
    %\State $C \gets C_\text{best}$ \Comment{Current circuit}
    \State $S \gets C_\text{best}$ \Comment{Search from all indices in the first step, then from modified indices from the previous step in the current iteration}
    \While{\textbf{true}}
        \State $found\_improvement \gets $ \textbf{false}
        \For{position $c \in S$} \label{alg:explore_greedy:position}
            \If{$t$ is applicable at $c$}
                \State $(C, S) \gets \Call{Apply}{C_\text{best}, c, t}$ \Comment{Returns new circuit and modified indices}
                \State $\Call{GreedyStep}{C, k-1, S}$
                \If{$found\_improvement$}
                    \State $S \gets \{c' \in C_\text{best} \mid c' \ge c\} \cup \{c' \in C_\text{best} \mid c' < c\}$ \label{alg:explore_greedy:wrap_around} \Comment{Wrap-around from current position}
                    \State \textbf{break}
                \EndIf
            \EndIf
        \EndFor
        \If{\textbf{not} $found\_improvement$}
            \State \textbf{break}
        \EndIf
    \EndWhile
\EndFor
\State \Return $C_\text{best}$
\EndFunction

\State

\Function{GreedyStep}{$C, k, S$}
\If{$k = 0$}
    \State $C \gets \Call{Exploit}{C}$ \Comment{Exploit for one step after exploring for $k$ steps}
    \If{$\Call{Cost}{C} < \Call{Cost}{C_{best}}$}
        \State $C_\text{best} \gets C$
        \State $found\_improvement \gets $ \textbf{true}
    \EndIf
    \State \Return
\EndIf
\For{position $c \in S$} \label{alg:explore_greedy:position}
    \For{$t \in \mathcal{T}$}
        \If{$t$ is applicable at $c$}
            \State $(C', S') \gets \Call{Apply}{C, c, t}$ \Comment{Returns new circuit and modified indices}
            \State $\Call{GreedyStep}{C', k - 1, S'}$ \label{alg:explore_greedy:recursive}
            \If{$found\_improvement$}
                \State \Return
            \EndIf
        \EndIf
    \EndFor
\EndFor
\EndFunction
\end{algorithmic}
\end{algorithm}

%% file: texts/5-eval.tex
\section{Evaluation}
\label{sec:eval}
%(for reference)
%In this section, we seek to answer the following questions: 1. What
% is the effect of POP on allocation quality and execution time on
% granular allocation problems? How does it compare to relevant
% heuristics? 2. Does POP work across a range of solvers and types of
% optimization problems? 3. How effective are POP’s client and
% resource splitting optimizations in generating high-quality
% allocations? 4. Howdoesrandompartitioning compare to other more
% sophisticated problem partitioning strategies?

% In this section, we seek to answer the following questions:

% \textbf{RQ1}: How does \optName compare with existing rule-based and
%     search-based optimizers?

% \textbf{RQ2} Why are we better than a vanilla combination (that ...)

% \mingkuan{Aidan: Figure for arithmetic mean gate reduction over time?
%   For GUOQ and Queso we can give time limit for 1min, 2min, 5min,
% 10min, 15min, 20min, 30min, 45min, 1h and connect the dots. 1 hour in total?}

\subsection{Experiment Setup}

To evaluate our framework, we employ a suite of quantum circuit
benchmarks consisting of 248 circuits compiled by
\guoq{}~\cite{xu2024optimizing, guoq-benchmarks}. This circuit set originally consisted of 250 circuits; however, \texttt{qft\_10} and
\texttt{qft\_16} were removed because they were identified as identity circuits. This suite aggregates diverse quantum circuits
from several established
frameworks~\cite{amy2014polynomial,hietala2021verified,nam2018automated,quartz-2022,li2024quarl,queso-2023,zulehner2018efficient}.
This benchmark suite includes variational circuits (QAOA, VQE),
simulation circuits (Trotterized TFIM, Heisenberg, and XY models),
classical arithmetic circuits (adders, GF multipliers, modular
arithmetic, and square-root circuits), and quantum algorithms
(Grover, Shor, QFT, and QPE).
We use gate count in the Nam gate set ($X, R_z, H, CX$) as our main
benchmark metric, following existing studies.
\Cref{fig:gate_scatter} shows the gate count distribution of the benchmark
suite.  The 248 circuits span five orders of magnitude in gate count.

\input{figtex/gate_scatter}

In the following experiments, we compare \optName against
\quartz~\cite{quartz-2022} (v0.3.2), \guoq~\cite{xu2024optimizing},
\queso~\cite{queso-2023}, %\qiskit~\cite{qiskit2024},
and
\voqc~\cite{hietala2021verified}.
% To maintain a consistent
% experimental environment, we perform all of our experiments in a
% modified version of the Docker container provided by Xu et
% al.~\cite{xu2024optimizing}.
% For \quartz, we evaluate the system both with and without its
% preprocessing pass.
% To leverage \quartz's internal decomposition logic, we provide it
% with circuits containing Toffoli ($CCZ$) gates when preprocessing is
% enabled. For all other baselines (and \quartz without preprocessing),
% we pre-decompose Toffoli gates into the Nam gate set using the
% standard fixed decomposition adopted by prior
% work~\cite{queso-2023,li2024quarl}.
%NOTE: Excluding \quartz preprocessing time from the one-hour budget gives
% that baseline an advantage not afforded to other tools. This should be
% clearly explained — ideally with a note on how long preprocessing takes —
% so readers can judge whether the comparison is still fair.
\input{figtex/eval_comparison_3x3.tex}

\input{figtex/averages_table}

We execute \guoq and \queso within the Docker environment provided by
their respective authors. For \queso, when optimizing for two-qubit
gates, we use the parameters \texttt{-g NAM -opt TWO\_Q --rules-dir
  /home/queso\_rules/
-search BEAM -temp 0 -q 8000 -resynth NONE}; when optimizing for
total gates, we use the parameters
\texttt{-g NAM -opt TOTAL --rules-dir /home/queso\_rules/ -search
BEAM -temp 0 -q 8000 -resynth NONE}, consistent with the authors'
recommendations.
For \guoq, when optimizing for two-qubit gates, we configure \guoq with
\texttt{-g NAM -opt TWO\_Q --rules-dir /home/queso\_rules/
--resynth-weight 180}; when optimizing for total gates, we configure \guoq with
\texttt{-g NAM -opt TOTAL --rules-dir /home/queso\_rules/
--resynth-weight 180}. \guoq{} only performs circuit resynthesis when
optimizing for two-qubit gates. Note that resynthesis is an
approximate optimization and does not preserve circuit semantics exactly.
All experiments were conducted on servers equipped
with AMD EPYC 7763 processors. Each optimizer is given one CPU core,
32 GiB of RAM, and a one-hour time limit for each run.

To capture the joint effect of $CX$ gates and single-qubit gates
on circuit quality, we estimate fidelity as the product of
per-gate fidelities~\cite{xu2024optimizing}, with $f_{1q} = 0.999$
for each single-qubit gate and $f_{2q} = 0.99$ for each two-qubit
gate, consistent with typical NISQ noise
models~\cite{murali2019noise, tannu2019not}:
\begin{equation}
  F = f_{1q}^{\, n_{1q}} \cdot f_{2q}^{\, n_{2q}},
  \label{eq:fidelity}
\end{equation}
where $n_{1q}$ and $n_{2q}$ denote the single- and two-qubit gate
counts, respectively.

\subsection{Main Results}
\input{figtex/eval_end_to_end.tex}

\label{sec:eval_per_circuit}
\Cref{fig:comparison_3x3} presents a per-circuit comparison against
\guoq{}, \queso{}, \voqc{}, and \quartz{} on output circuit fidelity, $CX$ gate
reduction, and total gate reduction. \optName{} achieves higher
fidelity than all four baselines
on the majority of circuits. \quartz{} was run without its
preprocessing pass, and we
evaluate this further in \Cref{sec:eval_quartz}.

The gate reduction rows of \Cref{fig:comparison_3x3}, together with
the averages in \Cref{tab:avg-comparison}, reveal distinct trade-off
profiles. Against \guoq{} in two-qubit gate optimization mode,
\optName{} achieves nearly identical $CX$ reduction ($24.61\%$ vs.
$25.41\%$) but dramatically better total gate reduction ($52.34\%$
vs. $-2.63\%$): \guoq{}'s resynthesis frequently introduces many
additional single-qubit gates, yielding negative total gate
reduction on many circuits, whereas \optName{} avoids this entirely.
Against \voqc{}, the situation is reversed: both achieve similar
total gate reductions ($52.34\%$ vs. $46.37\%$), but \optName{}
yields significantly greater $CX$ reduction ($24.61\%$ vs.
$4.24\%$). In terms of fidelity, \optName{} (average $0.3864$)
outperforms both \voqc{} ($0.3428$) and \guoq{} in two-qubit mode
($0.3475$), as it balances single- and two-qubit gate reduction
most effectively.

Against the remaining baselines, \queso{} (in both optimization
modes), \quartz{}, and \guoq{} in total gate optimization
mode, \optName{} achieves better results on all three metrics, as
shown in \Cref{tab:avg-comparison}.

\subsection{Optimization Efficiency}
In \Cref{sec:eval_per_circuit}, each optimizer is given a one-hour
time limit per circuit. To demonstrate \optName{}'s fast
convergence, we additionally evaluate it with only a one-minute
time limit per circuit. As shown in \Cref{tab:avg-comparison},
reducing the time budget by a factor of $60$ degrades \optName{}'s
average fidelity and gate reduction by less than $1\%$, yet it
still surpasses the one-hour fidelity and total gate reduction of every baseline.

\subsection{Quartz's Benchmark}
\label{sec:eval_quartz}
\quartz~\cite{quartz-2022} describes a preprocessing pass to
translate $CCZ$ gates into the Nam gate set ($X, R_z, H, CX$).
However, our main benchmark suite is already in the Nam gate set, and
it is hard to reconstruct the $CCZ$ gates to fully leverage
the Toffoli decomposition in \quartz's preprocessing pass. Therefore,
we also compare with \quartz{} on the 26 benchmark circuits used
by \quartz{}. This evaluation favors \quartz{} with preprocessing
because the preprocessing is run separately, and its time is not
counted towards the one-hour time limit.

\Cref{tab:eval_end_to_end} presents the gate counts achieved by each
optimizer on all 26 benchmark circuits. \optName{} achieves the lowest
or equally lowest
gate count on all of them except for one circuit compared to \quartz
with preprocessing. Compared to the purely rule-based \voqc{}, \optName{}
reduces gate count by an additional $7.3\%$ on average. Against the search-based
optimizers, \optName{} outperforms \quartz{} (without preprocessing)
by $17.8\%$,
\queso{} by $10.4\%$, \guoq{} by $5.1\%$, and \quartz{} (with
preprocessing) by $4.6\%$.  These
numbers are significant because they are
absolute percentages (relative to the original gate count instead of
optimized gate count), and each gate has a significant runtime and
fidelity cost on near-term quantum
computers. In addition, \quartz{} with preprocessing takes advantage of
higher-level knowledge about which gates originally compose a
Toffoli gate, yet \optName{} still matches or outperforms it on all
but one circuit. These results demonstrate that the interleaved
combination of rule-based and search-based optimization in
\optName{} yields consistently better circuit quality than either
approach alone, and typically subsumes the optimizations found by
\quartz{}'s preprocessing pass.

The following ablation studies use the same circuits as
\Cref{tab:eval_end_to_end}.

\subsection{Greedy Mode}
\label{sec:eval_greedy}

\begin{figure}[h]
  \centering
  \includegraphics[width=\linewidth]{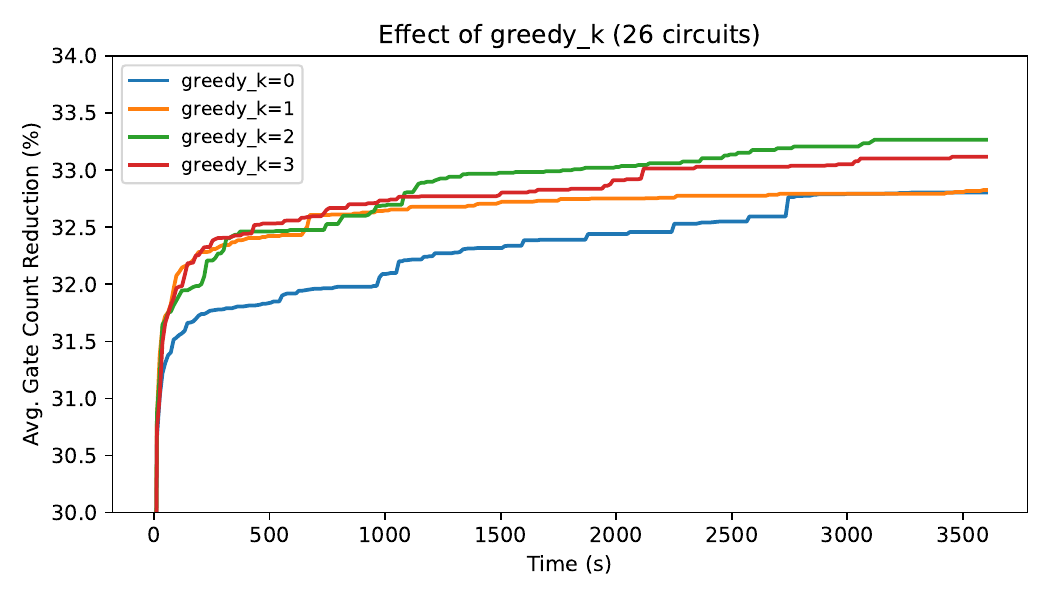}
  \caption{Effect of the greedy mode. Each curve shows \optName{} run
    with greedy mode (accepting only cost-reducing transformations)
    applied up to depth $k \le \mathrm{greedy\_k}$, with unrestricted
  search beyond.}
  \label{fig:greedy_k}
\end{figure}

\Cref{fig:greedy_k} shows the effect of
the greedy mode. While the final gate count reductions differ
by less than 0.5\% on average (32.80\% without greedy mode versus
33.27\% with greedy mode for $k\le 2$), the speed of convergence
varies substantially. Without the greedy mode, \optName{} takes 968
seconds to reach an average gate count reduction of 32\% on this
benchmark suite. Enabling the greedy mode for $k=1$ alone reaches
the same threshold in 89 seconds, an 11$\times$ speedup.

However, it is not always beneficial to extend greedy mode to larger search depths. When the greedy mode is only enabled for $k=1$, the
greedy phase takes 5.5 seconds on average. For $k \le 2$, it takes
82 seconds on average. Extending greedy mode to $k \le 3$ pushes the
greedy phase to 724 seconds on average, and two of the 26 circuits
(\texttt{gf2\^{}9\_mult} and \texttt{gf2\^{}10\_mult})
fail to finish the greedy phase within the one-hour budget.
The resulting final reduction (33.12\%) is also slightly below the
$k \le 2$ setting (33.27\%), since the extra time spent exhausting
$k=3$ greedy transformations crowds out the main search.
With a time limit of one hour, enabling greedy mode for $k \le 2$
strikes a balance for harvesting ``low-hanging fruit'' in circuit
optimization.

\subsection{Exploration Depth}
\label{sec:eval_k}
\begin{figure}[h]
  \centering
  \begin{subfigure}[t]{\linewidth}
    \includegraphics[width=\linewidth]{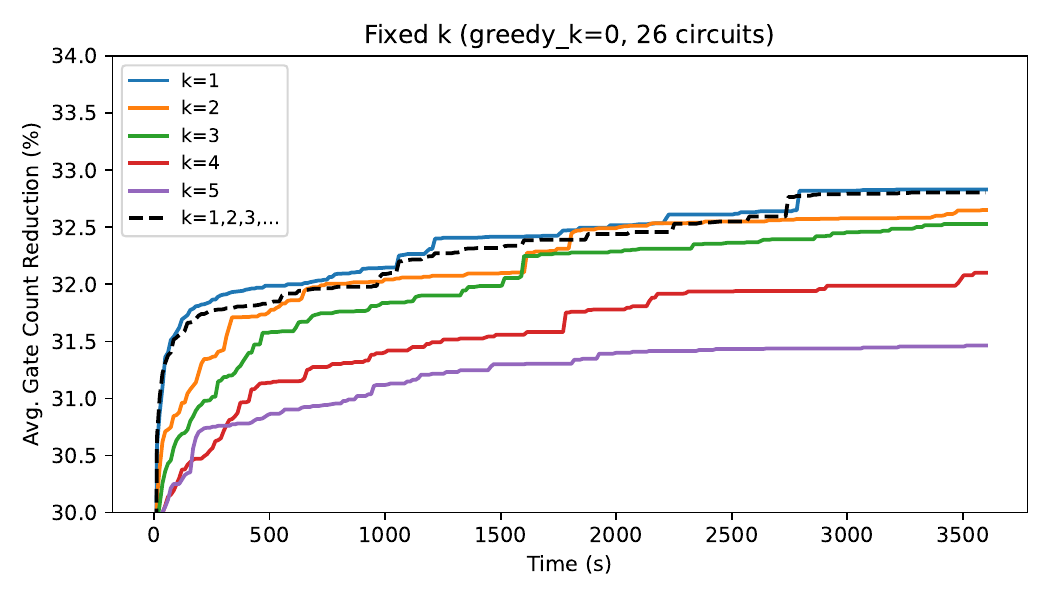}
    \caption{Fixed $k$ with greedy mode disabled.}
    \label{fig:single_k_gk0}
  \end{subfigure}
  \hfill
  \begin{subfigure}[t]{\linewidth}
    \includegraphics[width=\linewidth]{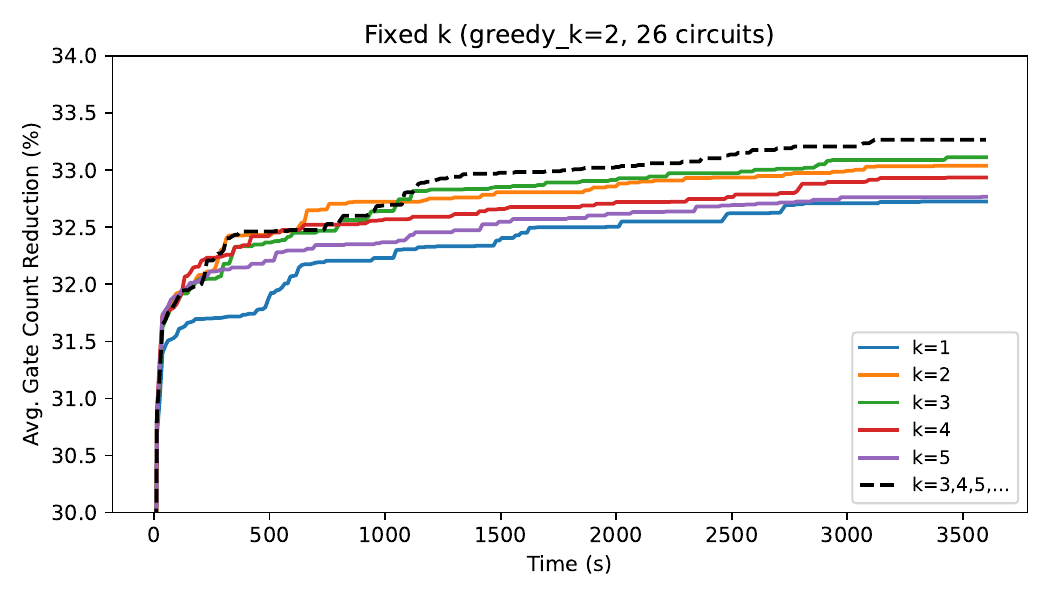}
    \caption{Fixed $k$ after greedy mode with $k=1,2$.}
    \label{fig:single_k_gk2}
  \end{subfigure}
  \caption{Effect of fixing the exploration depth $k$ in \optName{}.
    Each curve runs \optName{} with a single fixed $k\in\{1,\dots,5\}$
  (no depth advancement), averaged over 26 benchmark circuits.}
  \label{fig:single_k}
\end{figure}

\Cref{fig:single_k} shows the results of fixing the exploration
depth $k$ in \optName{}. Larger fixed $k$ explores deeper per
iteration but fewer iterations fit in the budget; small fixed $k$
iterates quickly but saturates early. Without the greedy
mode (\Cref{fig:single_k_gk0}), the final average reduction
decreases monotonically with $k$, from 32.83\% at $k=1$ down to
31.46\% at $k=5$, because there are still shallow optimizations left
unharvested, and deeper exploration incurs a higher per-iteration
cost, resulting in fewer iterations within the one-hour budget.

With the greedy mode enabled for $k \le 2$
and a fixed $k$ after that (\Cref{fig:single_k_gk2}), the ordering
shifts: $k=3$ becomes the
best single fixed depth (33.11\%), since the greedy phase already
harvests the shallow rewrites, and the remaining budget is spent on
deeper moves. Note that no single fixed $k$ matches the
advancing-$k$ baseline (33.27\%), and this motivates the
advancing-$k$ schedule used in \optName{}: it
exploits ``easier'' optimizations first with small $k$, then spends
the remaining budget on deeper exploration.

\subsection{Exploration Strategy}
\label{sec:eval_explore}

We now examine the impact of pool size $N_\text{pool}$ and branch
factor $N_\text{branch}$ on optimization quality.  As described in
\Cref{sec:greedy}, the greedy passes do not use
$N_\text{pool}$ and $N_\text{branch}$ parameters directly (equivalent to setting them to 1). For the main search
loop, we vary each
parameter independently over $\{1,2,3,4,5\}$ while holding the other
at its default ($N_\text{pool}=1$, $N_\text{branch}=3$),
with greedy mode enabled for $k \le 2$.

\begin{figure}[t]
  \centering
  \begin{subfigure}[t]{\linewidth}
    \includegraphics[width=\linewidth]{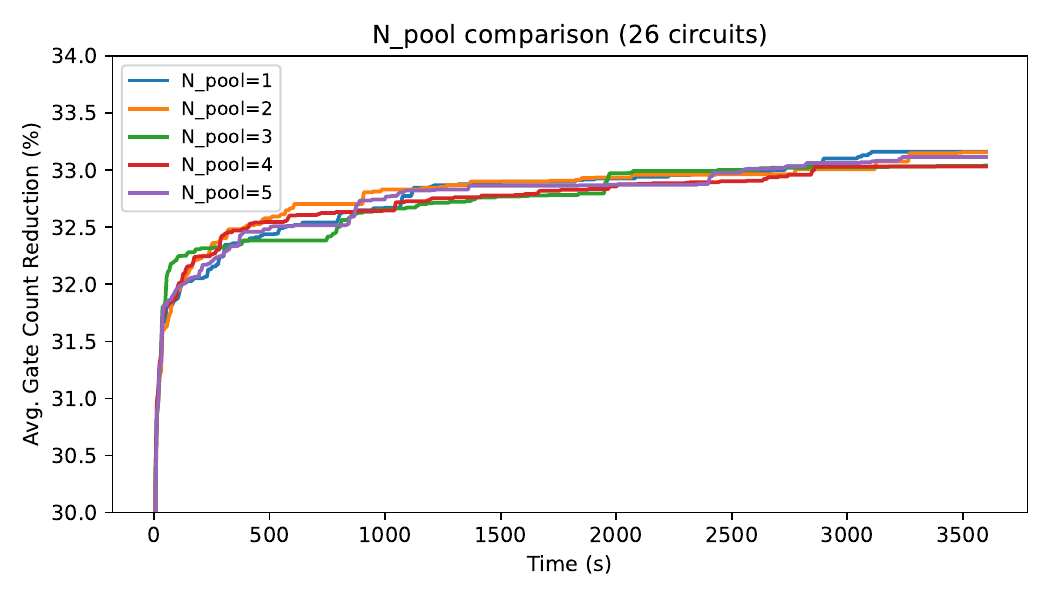}
    \caption{Pool size $N_\text{pool}$ ($N_\text{branch}=3$ fixed).}
    \label{fig:npool_comparison}
  \end{subfigure}
  \hfill
  \begin{subfigure}[t]{\linewidth}
    \includegraphics[width=\linewidth]{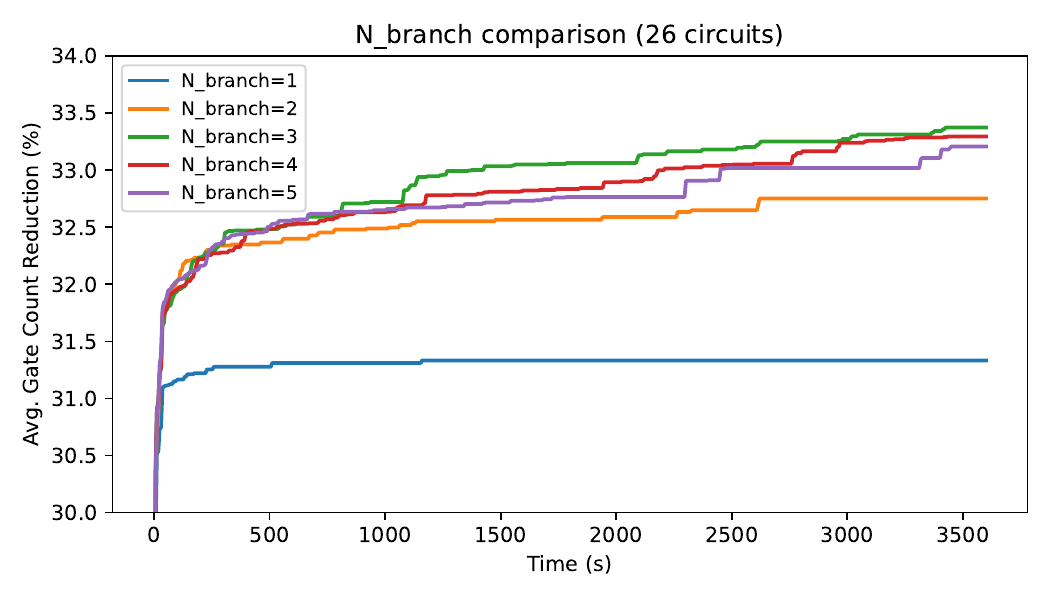}
    \caption{Branch factor $N_\text{branch}$ ($N_\text{pool}=1$ fixed).}
    \label{fig:nbranch_comparison}
  \end{subfigure}
  \caption{Effect of pool size $N_\text{pool}$ and branch factor
    $N_\text{branch}$ on average gate count reduction, over 26
    benchmark circuits with a one-hour timeout.  $N_\text{pool}$ has
    little effect on final quality, while $N_\text{branch}$ exhibits
  a sweet spot at $N_\text{branch}=3$.}
  \label{fig:exploration_ablation}
\end{figure}

\Cref{fig:exploration_ablation} shows the results.  Pool size
$N_\text{pool}$ has no meaningful effect on optimization quality:
across all five settings, the final average reduction ranges from
33.03\% to 33.16\% (a spread of 0.13\%), well within run-to-run
variance.  This insensitivity indicates that the cost-guided
best-first search already selects promising candidates from a
single-pool frontier, so widening the beam yields no additional
benefit within the one-hour budget.

Branch factor $N_\text{branch}$, in contrast, materially affects
final quality and exhibits a sweet spot at
$N_\text{branch}=3$. At $N_\text{branch}=1$, every expansion
retains only a single successor, so the search queue holds at most
one circuit at any time, the procedure collapses to a single
trajectory with no ability to recover from an
unfavorable local choice, and the average reduction drops to 31.33\%.
Raising $N_\text{branch}$ introduces the diversity needed to escape
such local minima, and quality climbs to 32.75\% at
$N_\text{branch}=2$ and peaks at 33.37\% at $N_\text{branch}=3$.
Beyond this point, the additional branches enlarge the frontier
faster than the one-hour budget can explore it, diluting the
per-branch search effort and slightly degrading quality (33.30\%
and 33.21\% at $N_\text{branch}=4$ and $5$).  We therefore use
$N_\text{pool}=1$ and $N_\text{branch}=3$ as the default
configuration for all other experiments.

%% file: figtex/gate_scatter.tex
\begin{figure}[t]
  \centering
  \includegraphics[width=\columnwidth]{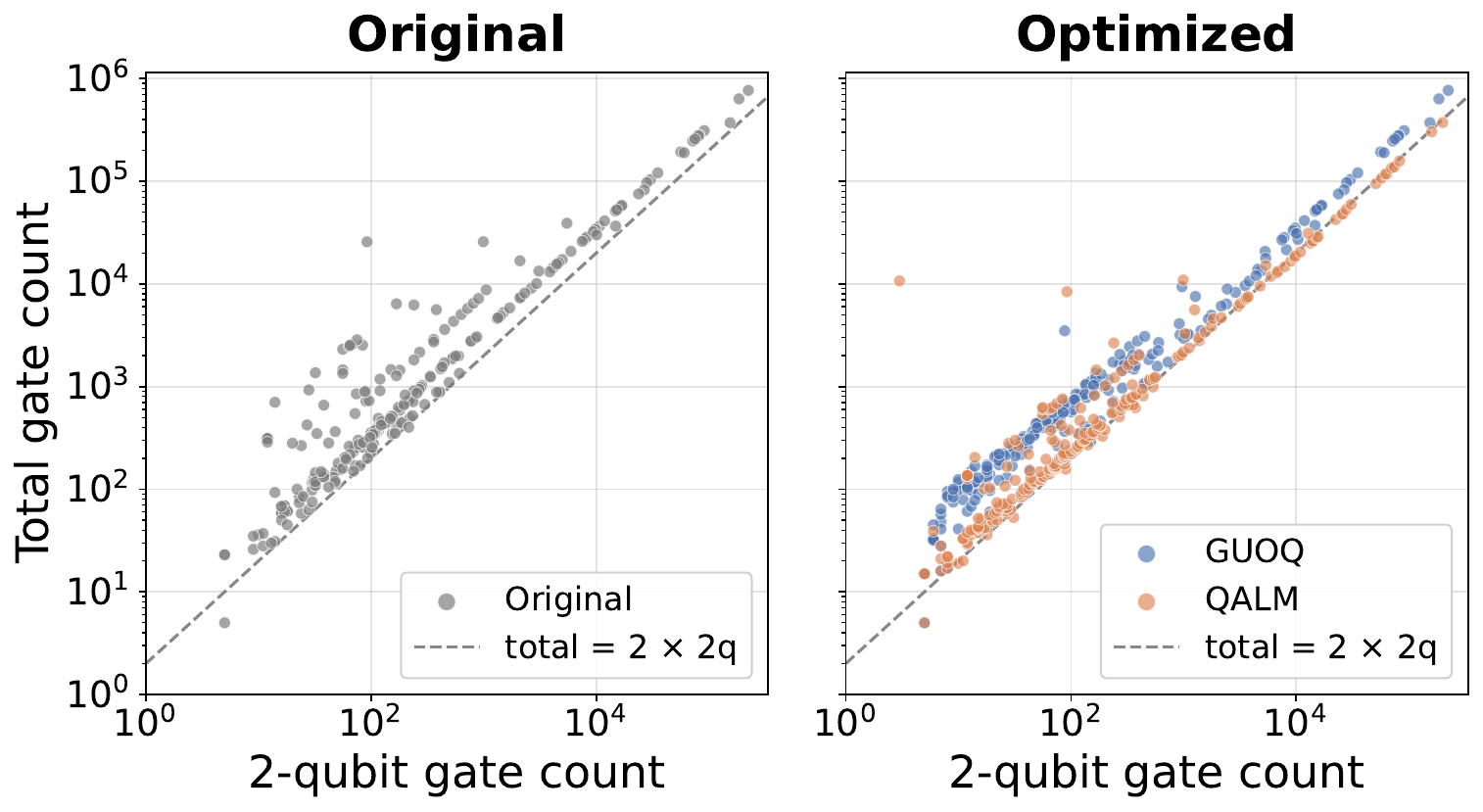}
  \caption{Gate count distribution in our benchmark. The dashed line
    indicates where the total gate count is twice the two-qubit gate
    count, or equivalently, where the number of single-qubit gates
    equals the number of two-qubit gates. The left panel shows the
    distribution before optimization, and the right panel shows the
    distribution after optimization using \optName{} and \guoq{} with
    two-qubit gate optimization mode, which is the strongest baseline
  in our evaluation.}
  \label{fig:gate_scatter}
\end{figure}

%% file: figtex/eval_comparison_3x3.tex
\begin{figure*}[t]
  \centering
  \includegraphics[width=\textwidth]{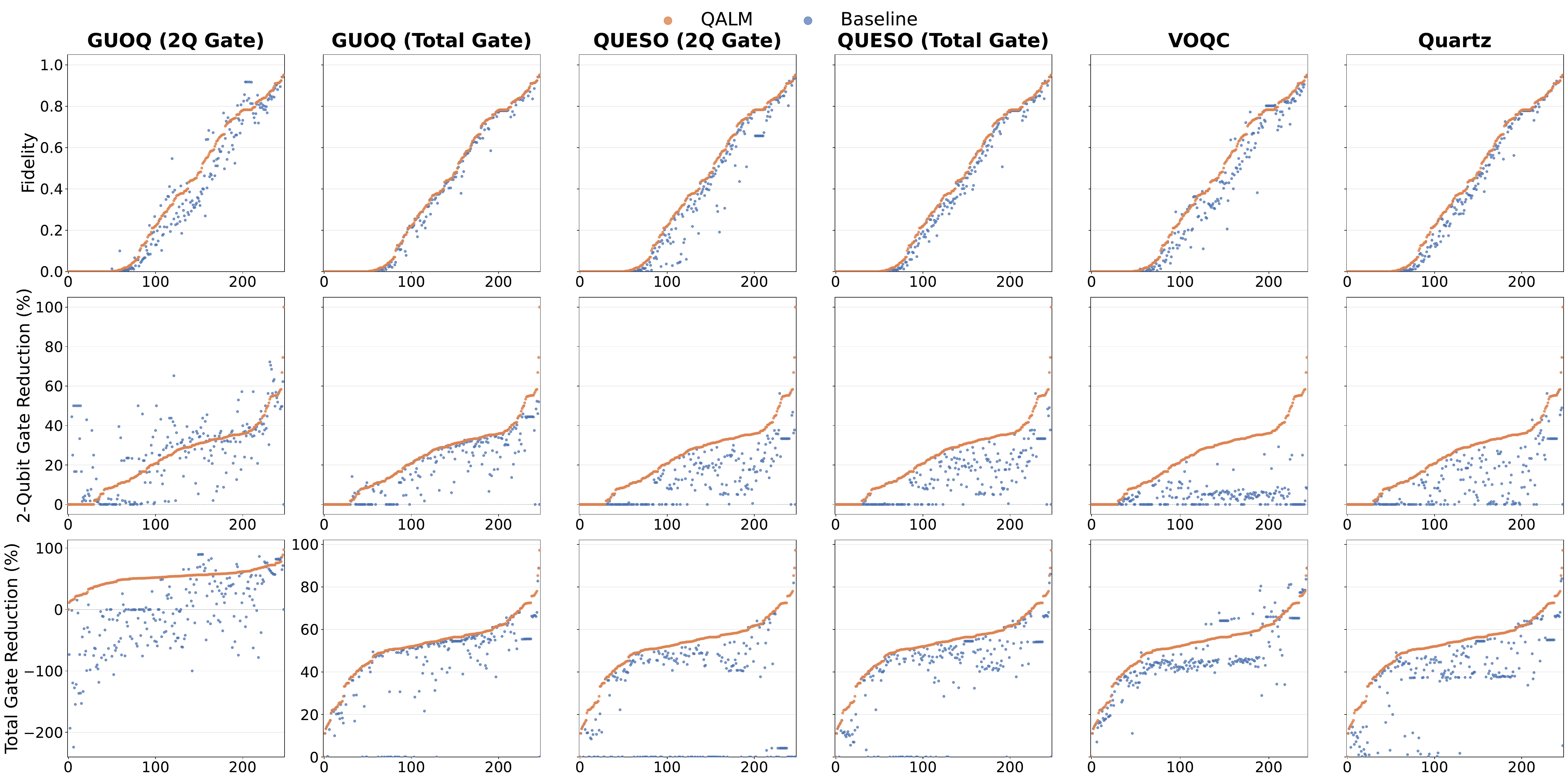}
  \caption{Per-circuit comparison of \optName{} against \guoq{},
    \queso{}, \quartz{}, and \voqc{}. Each column corresponds to a baseline
    optimizer, and each row shows a different metric: output circuit
    fidelity (top), $CX$ gate reduction percentage (middle), and total
    gate reduction percentage (bottom). Circuits are sorted by the
  \optName{} metric value. }
  \label{fig:comparison_3x3}
\end{figure*}

%% file: figtex/averages_table.tex
% Auto-generated by gen_averages_table.py — do not edit by hand.
% Averages of the dashed reference lines in the 3xN comparison figure.
\begin{table*}[t]
  \centering
%  \small
  \setlength{\tabcolsep}{4pt}
  \begin{tabular}{l c c c c c c c c }
    \toprule
%    Metric & QALM & GUOQ (2Q Gate) & GUOQ (Total Gate) & QUESO (2Q
%    Gate) & QUESO (Total Gate) & VOQC & Quartz \\
\multirow{2}{*}{Metric} & \multirow{2}{*}{QALM} & QALM & GUOQ & GUOQ & QUESO & QUESO & \multirow{2}{*}{VOQC} & \multirow{2}{*}{Quartz} \\
& & (1 minute) & (2Q Gate) & (Total Gate) & (2Q Gate) & (Total Gate) & & \\
    \midrule
    Fidelity                   & \textbf{0.3864} & 0.3747 & 0.3475 & 0.3676 & 0.3362 &
    0.3554 & 0.3428 & 0.3526 \\
    2Q Gate Reduction (\%)     & 24.61 & 21.18 & \textbf{25.41} & 18.83 & 13.19 &
    13.37 & 4.24 & 12.42 \\
    Total Gate Reduction (\%)  & \textbf{52.34} & 51.35 & -2.63 & 43.89 & 27.40 &
    39.72 & 46.37 & 40.73 \\
    \bottomrule
  \end{tabular}
  \caption{Per-metric averages across the benchmark suite. Each optimizer is given one hour per circuit, except for ``\optName{} (1 minute)'', which is given only one minute per circuit.}
  \label{tab:avg-comparison}
\end{table*}

%% file: figtex/eval_end_to_end.tex
\begin{table*}[ht]
\centering
\caption{Comparing \optName with existing optimizers on total gate count reduction under a 1-hour time limit, on 26 benchmark circuits used by \quartz. The best result for each circuit is in bold. \vspace{0.5em}}
\label{tab:eval_end_to_end}
\begin{tabular}{l|rrrrrr|r}
\hline
\multicolumn{1}{c|}{\multirow{2}{*}{\textbf{Circuit}}} & \multicolumn{1}{c}{\multirow{2}{*}{\textbf{Orig.}}} & \multicolumn{1}{c}{\multirow{2}{*}{\textbf{\voqc}}} & \multicolumn{1}{c}{\multirow{2}{*}{\textbf{\queso}}} & \multicolumn{1}{c}{\multirow{2}{*}{\textbf{\guoq}}} & \multicolumn{1}{c}{\textbf{\quartz}} & \multicolumn{1}{c|}{\textbf{\quartz}} & \multicolumn{1}{c}{\multirow{2}{*}{\textbf{\optName}}} \\
 &  &  &  &  & \multicolumn{1}{c}{\textbf{w/o Preprocess}} & \multicolumn{1}{c|}{\textbf{w/ Preprocess}} &  \\
\hline
\texttt{adder\_8} & 900 & 596 & 645 & 578 & 897 & 621 & \textbf{512} \\
\texttt{barenco\_tof\_3} & 58 & 40 & \textbf{38} & 39 & \textbf{38} & \textbf{38} & \textbf{38} \\
\texttt{barenco\_tof\_4} & 114 & 72 & 68 & 72 & 90 & 72 & \textbf{62} \\
\texttt{barenco\_tof\_5} & 170 & 104 & 98 & 102 & 134 & 104 & \textbf{94} \\
\texttt{barenco\_tof\_10} & 450 & 264 & 248 & 262 & 420 & 264 & \textbf{236} \\
\texttt{csla\_mux\_3} & 170 & 158 & 148 & 145 & 146 & 148 & \textbf{144} \\
\texttt{csum\_mux\_9} & 420 & 280 & 392 & 346 & 400 & \textbf{279} & 280 \\
\texttt{gf2\^{}4\_mult} & 225 & 186 & 195 & 169 & 199 & 173 & \textbf{167} \\
\texttt{gf2\^{}5\_mult} & 347 & 287 & 306 & 271 & 316 & 276 & \textbf{265} \\
\texttt{gf2\^{}6\_mult} & 495 & 401 & 426 & 376 & 462 & 383 & \textbf{370} \\
\texttt{gf2\^{}7\_mult} & 669 & 543 & 603 & \textbf{500} & 648 & 524 & \textbf{500} \\
\texttt{gf2\^{}8\_mult} & 883 & 706 & 815 & 673 & 880 & 695 & \textbf{660} \\
\texttt{gf2\^{}9\_mult} & 1095 & 879 & 996 & 824 & 1079 & 861 & \textbf{805} \\
\texttt{gf2\^{}10\_mult} & 1347 & 1065 & 1225 & 1046 & 1327 & 1040 & \textbf{972} \\
\texttt{mod5\_4} & 63 & 51 & 33 & 26 & \textbf{24} & \textbf{24} & \textbf{24} \\
\texttt{mod\_mult\_55} & 119 & 92 & 100 & 101 & 98 & 93 & \textbf{86} \\
\texttt{mod\_red\_21} & 278 & 184 & 198 & 198 & 215 & 202 & \textbf{175} \\
\texttt{qcla\_adder\_10} & 521 & 416 & 452 & 395 & 488 & 401 & \textbf{385} \\
\texttt{qcla\_com\_7} & 443 & 269 & 331 & 289 & 418 & 288 & \textbf{252} \\
\texttt{qcla\_mod\_7} & 884 & 678 & 791 & 657 & 878 & 646 & \textbf{622} \\
\texttt{rc\_adder\_6} & 200 & \textbf{152} & 176 & 180 & 180 & \textbf{152} & \textbf{152} \\
\texttt{tof\_3} & 45 & \textbf{35} & \textbf{35} & \textbf{35} & \textbf{35} & \textbf{35} & \textbf{35} \\
\texttt{tof\_4} & 75 & 55 & 55 & 55 & 55 & 55 & \textbf{51} \\
\texttt{tof\_5} & 105 & 75 & 75 & 75 & 76 & 75 & \textbf{71} \\
\texttt{tof\_10} & 255 & 175 & 175 & 180 & 177 & 175 & \textbf{165} \\
\texttt{vbe\_adder\_3} & 150 & 89 & 83 & 80 & 114 & 89 & \textbf{71} \\
\hline
\textbf{Geo. Mean Reduction} & -- & 27.0\% & 23.9\% & 29.2\% & 16.5\% & 29.7\% & 34.3\% \\
\hline
\end{tabular}
\end{table*}

%% file: texts/6-related_works.tex
\section{Related Work}
\label{sec:related}

This section surveys rule-based and search-based optimization
methods, along with prior efforts to combine them.

\subsection{Rule-Based Optimization}

Rule-based optimizers apply predefined circuit transformation rules
in a manner analogous to peephole optimization in classical
compilers. Representative approaches include Nam et
al.~\cite{nam2018automated}, which systematically applies rotation
merging and gate cancellation rules, and
VOQC~\cite{hietala2021verified}, which provides formally verified
implementations of these rules. PyZX~\cite{kissinger2020Pyzx} takes a
different approach by converting circuits to ZX-diagrams and applying
ZX-calculus rewrite rules, though extracting optimized gate-based
circuits from ZX-diagrams is computationally
expensive~\cite{de2022circuit}. Phase polynomial
methods~\cite{amy2014polynomial,amy2019formal,chen2025phasepoly}
offer another rule-based strategy by representing circuits
algebraically for rotation gate merging. 
OAC~\cite{arora2025local} and POPQC~\cite{liu2025popqc} apply external optimizers to circuit segments with provable local optimality guarantees and improved speed. Their inherently greedy nature makes them more akin to rule-based methods.

These methods are relatively efficient and scalable, but their greedy nature
limits optimization quality.

\subsection{Search-Based Optimization}

Search-based optimizers explore the space of equivalent circuits by
tolerating cost-increasing transformations.
\quartz~\cite{quartz-2022} and \queso~\cite{queso-2023} automatically
generate equivalence rules and use circuit cost to guide the
exploration of circuit variants.
QSearch~\cite{qsearch} and QFast~\cite{qfast} use numerical
optimization to synthesize circuit blocks, also supporting
approximate optimization.
These methods achieve better optimization quality but face
scalability challenges due to exponential search space growth.

\subsection{Hybrid Optimization Strategies}

Several recent works have attempted to bridge the gap between
rule-based efficiency and search-based quality. Reinforcement
learning approaches~\cite{li2024quarl,fosel2021quantum} train agents
to select which optimization actions to apply, learning implicit
strategies for balancing exploration and exploitation.
\guoq~\cite{xu2024optimizing} introduces a ``fast-slow'' framework
that combines search-based and resynthesis methods, though its ``fast''
phase still relies on search-based exploration rather than pure rule
application, and it primarily targets approximate optimization.

These hybrid approaches represent important progress, but they either
still depend heavily on search-based exploration in their fast paths,
or require expensive training procedures. In contrast, \optName
achieves efficient optimization by strategically interleaving
lightweight rule-based passes with targeted search-based exploration,
avoiding local minima while maintaining competitive runtime.

\subsection{Optimization Objectives}

While this paper focuses on gate count reduction, quantum circuit
optimization encompasses diverse objectives depending on the target
platform. In the NISQ era, circuit fidelity is critical, motivating
noise-aware optimizations~\cite{murali2019noise,tannu2019not} and
hardware-specific compilation for topology, gate set, and pulse
characteristics~\cite{molavi2022qubit,lye2015determining,itoko2020optimization,li2019tackling,nottingham2023decomposing,wu2021tilt,shi2019optimized,gokhale2020optimized,davis2020towards}.
For fault-tolerant quantum computing, the $T$ gate dominates resource costs, making $T$ count the primary optimization target. Techniques
such as phase folding~\cite{amy2014polynomial,amy2019formal} and the
Feynman toolkit~\cite{amy2019formal} efficiently reduce $T$ count
while also optimizing other metrics like $CX$ count.

%% file: texts/7-conclusion.tex
\section{Conclusion and Outlook}
We identified a fundamental bottleneck in search-based quantum
circuit optimization: once a truly promising point has a higher cost
than the current local minimum, a pure search must expend
exponentially many steps to verify that the point actually leads to
a better basin. This inability to cheaply distinguish genuinely
promising points from cost-increasing detours is what forces the
long-standing compromise between speed and quality.

We presented \optName{}, which resolves this bottleneck by
interleaving search-based exploration with rule-based
exploitation in a single control loop. Each exploit phase
immediately descends from every candidate uncovered by the
preceding explore phase, so promising points are verified in a
single rule-based pass rather than through exponential search. By
progressively advancing the exploration depth $k$, \optName{}
exhausts shallow optimizations first and reserves its budget for
the deeper moves that matter.

Our evaluation confirms that this design eliminates the historical
trade-off between speed and quality: \optName{} outperforms existing
rule-based and search-based optimizers on total gate count reduction,
and it surpasses the one-hour results of all
search-based baselines within the first minute.

Future directions include extending \optName{} to optimize for other
metrics such as
$T$ count, fidelity, and circuit depth, as well as investigating
circuit-specific parameter scheduling to further enhance search efficiency. It would be interesting to see how these results would compare with GUOQ's optimization objectives that we did not run, such as fidelity optimization.

%% file: texts/8-ack.tex
\section*{Acknowledgments}

Claude and Gemini were used to generate scripts for running the experiments,
making plots, and proofreading the text.

% \section*{Availability}
% %-------------------------------------------------------------------------------

% USENIX program committees give extra points to submissions that are
% backed by artifacts that are publicly available. If you made your code
% or data available, it's worth mentioning this fact in a dedicated
% section.